\documentclass[twocolumn, showpacs, aps, eqsecnum]{revtex4}

\usepackage{graphicx}%
\usepackage{dcolumn}
\usepackage{amsmath}

\begin{document}


\title[Short Title]{Control of fine-structure splitting and excitonic binding energies in selected individual InAs/GaAs quantum dots}

\author{R. Seguin}
	\email{seguin@sol.physik.tu-berlin.de}
\author{A. Schliwa}
\author{T. D. Germann}
\author{S. Rodt}
\author{K. P\"otschke}
\author{A. Strittmatter}
\author{U. W. Pohl}
\author{D. Bimberg}
 
\affiliation{
Institut f\"ur Festk\"orperphysik, Technische Universit\"at Berlin, D-10623 Berlin, Germany}

\author{M. Winkelnkemper}
\affiliation{
Institut f\"ur Festk\"orperphysik, Technische Universit\"at Berlin, D-10623 Berlin, Germany}
\affiliation{Fritz-Haber-Institut der Max-Planck-Gesellschaft, D-14195 Berlin, Germany}

\author{T. Hammerschmidt}
\affiliation{Fritz-Haber-Institut der Max-Planck-Gesellschaft, D-14195 Berlin, Germany}

\author{P. Kratzer}
\affiliation{Fritz-Haber-Institut der Max-Planck-Gesellschaft, D-14195 Berlin, Germany}
\affiliation{Fachbereich Physik, Universit\"at Duisburg-Essen, D-47048 Duisburg, Germany}

\date{\today}

\begin{abstract}
A systematic study of the impact of annealing on the electronic properties of single InAs/GaAs quantum dots (QDs) is presented. Single QD cathodoluminescence spectra are recorded to trace the evolution of one and the same QD over several steps of annealing. A substantial reduction of the excitonic fine-structure splitting upon annealing is observed. In addition, the binding energies of different excitonic complexes change dramatically. The results are compared to model calculations within 8-band {\bf k}$\cdot${\bf p} theory and the configuration interaction method, suggesting a change of electron and hole wave function shape and relative position.
\end{abstract}


\maketitle
\newpage

Single quantum dots (QDs) are building blocks for numerous modern devices including single-photon emitters \cite{benson} and storage devices \cite{geller, cortez}. Targeted tailoring of their electronic properties is of utmost importance for device functionality. A vanishing exciton fine-structure splitting (FSS) of QDs, for example, is the key parameter for generating entangled photon pairs \cite{santori, stevenson02}. FSS post-growth modification was recently demonstrated using external electric \cite{kowalik} and magnetic fields \cite{stevenson} as well as externally applied stress \cite{seidl}. In order to limit the complexity of a final device, it is desirable to modify the FSS permanently, e.g., by precise variation of the QD structure. Annealing can considerably alter the electronic structure of QD ensembles leading to a change of FSS \cite{young, langbein, tarta} and biexciton binding energy \cite{young}. While Refs.~\onlinecite{langbein} and \onlinecite{tarta} measured average properties of QD ensembles, Young et al.~\cite{young} determined the electronic properties of individual QDs by performing single QD spectroscopy. However, they analyzed QDs randomly chosen before and after annealing. In contrast to that, in this letter a systematic study of the influence of annealing on the emission characteristics of {\it one and the same} QD for consecutive steps of annealing is presented. Fine-structure splittings and excitonic binding energies are determined. By comparing the experimental results to calculations within the framework of 8-band {\bf k}$\cdot${\bf p} theory and the configuration interaction (CI) method \cite{stier99}, the impact of annealing on size, position and orientation of the participating wave functions is discussed.

\begin{figure} 
  \includegraphics[width=.99\columnwidth]{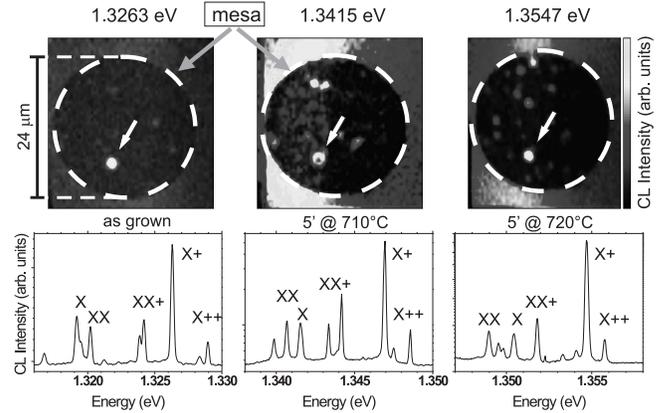}
  \caption{In the upper part, monochromatic CL images of the mesa are shown for the as-grown case and after two consecutive annealing steps. The position of the examined QD is indicated by a white arrow. The lower part shows CL spectra of the QD before and after the two respective annealing steps. Lines originating from neutral and charged excitonic complexes are identified.}
\label{figure1}\end{figure}

The InAs QDs were grown in an AIXTRON 200/4 metal-organic chemical vapor deposition (MOCVD) reactor in GaAs matrix on GaAs(001) substrates. A 300~nm thick GaAs buffer layer followed by a 60 nm Al$_{0.6}$Ga$_{0.4}$As diffusion barrier and 90~nm GaAs were grown. For the QDs nominally 1.9 monolayers of InAs were deposited followed by a 540~s growth interruption.  Subsequently, the QDs were capped with 50~nm GaAs. Finally, a 20~nm Al$_{0.33}$Ga$_{0.67}$As diffusion barrier and a 10~nm GaAs capping layer were deposited. During the growth interruption, the QDs undergo a ripening process, i.e., material is transported from small to large QDs, leading to a redshift of the ensemble luminescence \cite{potschke}. Here the ensemble peak is centered at 1.06~eV at T=10~K. While most small QDs dissolve during the long growth interruption, some remain, leading to an ultralow QD density ($<10^7$ per cm$^2$) in the 1.25-1.35~eV spectral range. 

The two consecutive annealing steps lasted five minutes at 710~$^{\circ}$C and 720~$^{\circ}$C, respectively, performed under As atmosphere in the MOCVD reactor in order to stabilize the sample surface.

The sample was examined using a JEOL JSM 840 scanning electron microscope equipped with a cathodoluminescence (CL) setup. The luminescence was dispersed by a 0.3~m monochromator equipped with a 1200 lines/mm grating. The light was detected using a liquid-nitrogen cooled Si charge-coupled-device camera. The resolution of the setup is $\approx$140~$\mu$eV. Using line shape analysis, the energetic position of single lines could be determined withtin an accuracy better than 20~$\mu$eV. The sample was mounted onto the tip of a helium-flow cryostat providing temperatures as low as 6 K used for all experiments throughout this letter.

Circular mesas of 24~$\mu$m in diameter were etched into the sample surface. Figure \ref{figure1} shows monochromatic CL images of such a mesa viewed from the top. One can clearly see, that the QD density is low enough to identify individual QDs and their position within the mesa. It is thus possible to relocate one specific QD after the sample has undergone an annealing step.  

\begin{figure}
\includegraphics[width=.99\columnwidth]{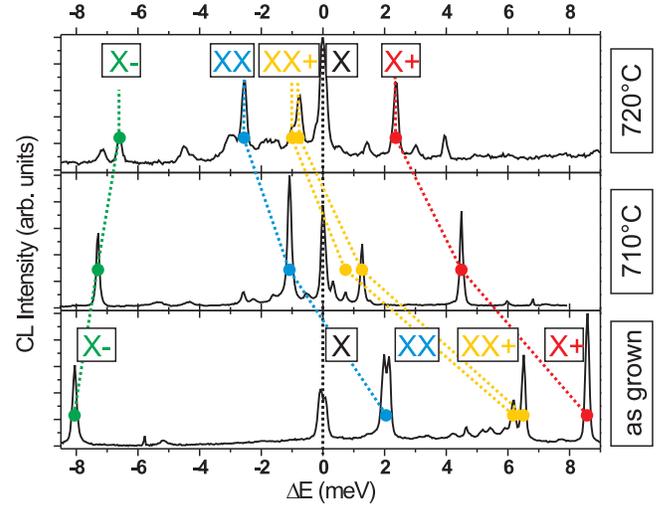}
  \caption{Effect of two consecutive annealing steps on the spectrum of a single QD. 0 meV corresponds to the respective neutral exciton recombination energy (1.2738~eV for as-grown, 1.3002 for 710~$^{\circ}$C, and 1.3174 eV for 720~$^{\circ}$C). The X$^{-}$-line shifts to higher energies with respect to the X-line (i.e.\ the X$^{-}$ becomes less binding). All other lines follow the opposite trend (i.e.\ become more binding).}
\label{figure2}\end{figure}

The QD spectra change dramatically due to a modification of the QD structure during the annealing procedure. Figure \ref{figure2} shows the effect of the annealing steps on the spectrum of a particular QD. Neutral excitons (X), biexcitons (XX) and charged [positively (X$^+$), negatively (X$^-$)] excitons could be identified following Ref.~\onlinecite{rodt05}. For the QD shown in Fig.~\ref{figure2}, the X transition energy increased by 43.6~meV after both annealing steps . For ease of comparison of the annealing induced variations, the energetic position of the X line has been set to 0~meV. The XX shifts to lower energies with respect to the X line, changing its character from anti-binding \mbox{($E^B_{XX}=-2.1$ meV)} to binding ($E^B_{XX}=2.6$~meV) with a total change in binding energy $\Delta E_{XX}^{B} = 4.7$~meV. Likewise, the X$^+$ binding energy increases by \mbox{$\Delta E_{X^{+}}^{B}=$ 6.3~meV}. The X$^-$, on the other hand, shows the opposite trend, becoming less binding with its binding energy decreasing by \mbox{$\Delta E_{X^{-}}^{B}=$ -1.3~meV}. The annealing process thus has a surprisingly large impact on all binding energies, suggesting a drastic change of the involved wave functions and/or their mutual position.

\begin{figure}
\includegraphics[width=.99\columnwidth]{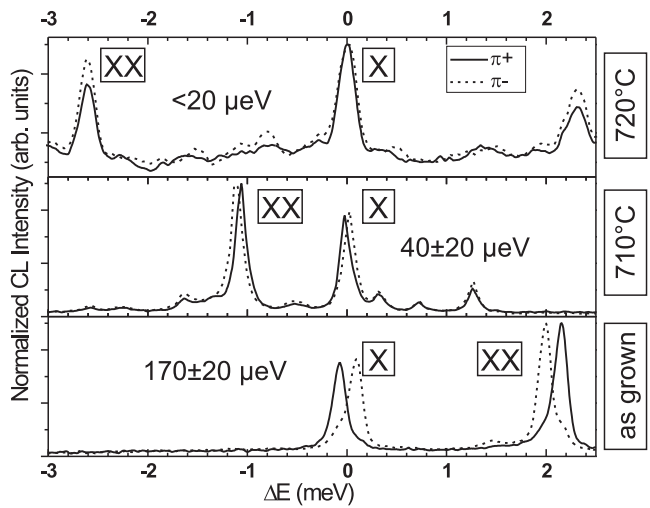}
  \caption{Polarization dependent measurements of the same QD depicted in Fig.~\ref{figure2} reveal the excitonic fine-structure splitting. It decreases from $170 \pm 20$~${\mu}$eV to less than 20~${\mu}$eV after the second annealing step.}
\label{figure3}\end{figure}

Additionally, the excitonic FSS was recorded by performing polarization dependent measurements (Fig.\ \ref{figure3}). For this particular QD it decreased from 170~$\mu$eV to less than 20~$\mu$eV. It should thus be possible to reduce the FSS in a controlled way below the homogeneous linewidth of the X-line (which is on the order of a few $\mu$eV at liquid He temperatures), a prerequisite for the generation of polarization-entangled photon pairs \cite{santori, stevenson02}. The general trend of decreasing the FSS \cite{young, langbein, tarta} and increasing the XX binding energy \cite{young} by annealing has also been observed by other authors. However, only by measuring the same QD before and after annealing a detailed analysis of the interplay of QD morphology, wave function position and shape, and excitonic properties can be conducted. While in this letter we only show results for one specific QD, the described observations and trends are typical for all analyzed QDs.

The experimental results were modeled using 8-band {\bf k}$\cdot${\bf p} theory for the single particle orbitals and the CI method for the few-particle states. The as-grown QD was assumed to have truncated pyramidal shape with \{101\} side facets, a height of 1.42~nm, a lateral size of 11.3~nm, and an In content of 100~\%. There is some uncertainty concerning these numbers since the determination of the structure of a few small QDs within an ensemble of large QDs is very difficult. While reliable information about the structure of similarly grown QDs exists \cite{potschke}, the influence of the long growth interruption on the morphology of the QDs is unknown. Our model QD yields an X transition energy of 1.08~eV. In our experiments it is likely that a slight fraction of Ga is incorporated into the QDs during growth, leading to a higher emission energy than for the model QD. The annealing process and the resulting exchange of In and Ga atoms grade the interfaces between matrix material and QD. This effect was simulated by applying a smoothing algorithm for each annealing step, corresponding to Fickian diffusion. The resulting atomistic structures were relaxed with a recently developed bond-order potential of the Abell-Tersoff type that is particularly suited for InAs/GaAs heterostructures \cite{hammerschmidt}. To be more precise, we performed slab calculations with periodic boundary conditions in the lateral directions, a conjugate-gradient scheme to minimize the total energy until the maximum force on an atom in the system was below 1 meV/\AA, and the scheme outlined in Ref.~\cite{pryor} to determine the atomistic strain tensor. The single-particle orbitals were then derived from a strain-dependent 8-band {\bf k}$\cdot${\bf p} Hamiltonian accounting for band coupling and band mixing. The piezoelectric field was computed using the first and second order piezoelectric tensor \cite{bester_a, bester_b}. The excitonic states were calculated using the CI method by expanding the respective excitonic Hamiltonian into a basis of antisymmetrized products of single particle wave functions built from a total of six electron and six hole wave functions. This procedure accounts for direct Coulomb interaction, exchange, and in part correlation \cite{rodt05}.

\begin{figure}
\includegraphics[width=.99\columnwidth]{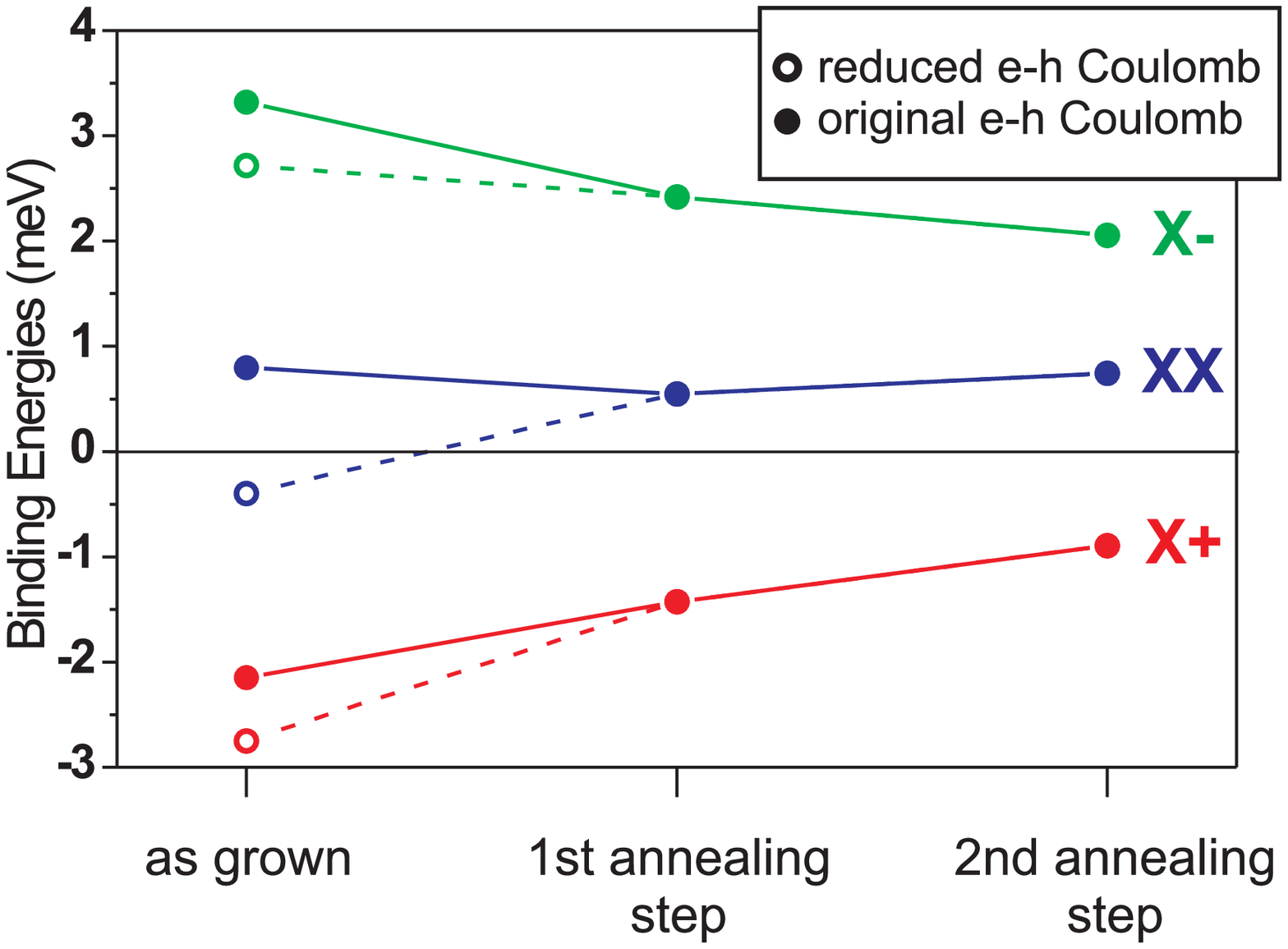}
  \caption{Calculated binding energies for X$^-$, X$^+$, and XX. Open symbols correspond to a reduction of electron-hole Coulomb interaction by 0.6 meV simulating a misalignment of electron and hole wave functions (see text).}
\label{figure4}\end{figure}

Results for the binding energies of X$^+$, X$^-$, and XX are shown in Fig.~\ref{figure4}. The experimentally observed energetic order of X$^-$ and X$^+$ is well reproduced by our theory. Also, the general trend of the binding energies upon annealing (X$^-$ becomes less binding, X$^+$ becomes more binding) is in good agreement. However, there are some discrepancies between theory and experiment: The change of character of the biexciton (i.e.~from anti-binding to binding) as observed in the experiment is not reproduced by the modeling. Also, the experiment revealed a much stronger relative shift for X$^+$ and XX than for X$^-$ (see Fig.~\ref{figure2}).

These discrepancies imply that an important effect is not correctly modeled. Remarkably the disagreements can be resolved, when the attractive Coulomb interaction between electron and hole ($J_{eh}$) is artificially lowered for the as-grown QD by 0.6 meV, corresponding to a reduction of 3\%, while repulsive electron-electron ($J_{ee}$) and hole-hole ($J_{hh}$) Coulomb terms are left unaltered. Such an effect may result from a reduced electron-hole wave function overlap caused by either a slight elongation and misalignment of the wave functions or a mutual vertical displacement. 

Using the assumption of reduced $J_{eh}$ for the as-grown QD we observe both the biexciton crossover as well as the relative insensitivity of X$^-$ binding energy to annealing as seen in the experiment (open symbols in Fig.~\ref{figure4}). A symmetrization of the wave functions with annealing is further supported by the drastic reduction of FSS \cite{langbein}. 

In conclusion, we have recorded emission spectra of single QDs and followed their evolution under an annealing procedure. The binding energies of different excitonic complexes change on the order of several meV and the FSS decreases from 170 $\mu$eV to less than 20~$\mu$eV. These results can be understood by a change of electron and hole wave function shape and mutual position. We have thus demonstrated a powerful tool to tailor single QDs' electronic properties for their use in potential applications. In particular, zero FSS, essential for emission of entangled photon pairs, can be enforced.

The calculations were performed on the IBM pSeries 690 supercomputer at HLRN within project No.~bep00014. This work was supported by the Deutsche Forschungsgemeinschft in the framework of SfB 296 and the SANDiE Network of Excellence of the European Commision, Contract No. NMP4-CT-2004-500101.


\newpage
\begin{thebibliography}{9}

\bibitem{benson}
O. Benson, C. Santori, M. Pelton, and Y. Yamamoto, Phys. Rev. Lett. {\bf 84}, 2513 (2000).

\bibitem{cortez}
S. Cortez, O. Krebs, S. Laurent, M. Senes, X. Marie, P. Voisin, R. Ferreira, G. Bastard, J.-M. G\'erard, and T. Amand, Phys. Rev. Lett. {\bf 89}, 207401 (2002).

\bibitem{geller}
A. Marent, M. Geller, D. Bimberg, A.~P. Vasi'ev, E.~S. Semenova, A.~E. Zhukov, and V.~M. Ustinov, Appl. Phys. Lett. {\bf 89}, 072103 (2006).

\bibitem{santori}
C. Santori, D. Fattal, M. Pelton, G.~S. Solomon, and Y. Yamamoto, Phys. Rev. B {\bf 66}, 045308 (2002).

\bibitem{stevenson02}
R.~M. Stevenson, R.~M. Thompson, A.~J. Shields, I. Farrer, B.~E. Kardynal, D.~A. Ritchie, and M. Pepper, Phys. Rev. B {\bf 66}, 081302(R) (2002).

\bibitem{kowalik}
K. Kowalik, O. Krebs, A. Lema{\^i}tre, S. Laurent, P. Senellart, P. Voisin, and J.~A. Gaj, Appl. Phys. Lett. {\bf 86}, 041907 (2005).

\bibitem{stevenson}
R.~M. Stevenson, R.~J. Young, P. See, D.~G. Gevaux, K. Cooper, P. Atkinson, I. Farrer, D.~A. Ritchie, and A. Shields, Phys. Rev. B {\bf 73}, 033306 (2006).

\bibitem{seidl}
S. Seidl, M. Kroner, A. H\"ogele, K. Karrai, R.~J. Warburton, A. Badolato, and P.~M. Petroff, Appl. Phys. Lett. {\bf 88}, 203113 (2006).

\bibitem{langbein}
W. Langbein, P. Borri, U. Woggon, V. Stavarache, D. Reuter, and A.~D. Wieck, Phys. Rev. B {\bf 69}, 161301(R) (2004).

\bibitem{tarta}
A.~I. Tartakovskii, M.~N. Makhonin, I.~R. Sellers, J. Cahill, A.~D. Andreev, D.~M. Whittaker, J.-P.~R. Wells, A.~M. Fox, D.~J. Mowbray, M.~S. Skolnick, K.~M. groom, M.~J. Steer, H.~Y. Liu, and M. Hopkinson, Phys. Rev. B {\bf 70}, 193303 (2004).

\bibitem{young}
R.~J. Young, R.~M. Stevenson, A.~J. Shields, P. Atkinson, K. Cooper, D.~A. Ritchie, K.~M.Groom, A.~I. Tartakovskii, and M.~S. Skolnick, Phys. Rev. B {\bf 72}, 113305 (2005).

\bibitem{stier99} 
O. Stier, M. Grundmann, and D. Bimberg, Phys. Rev. B \textbf{59}, 5688 (1999).

\bibitem{potschke}
U.~W. Pohl, K. P\"otschke, A. Schliwa, F. Guffarth, D. Bimberg, N.~D. Zakharov, P. Werner, M.~B. Lifshits, V.~A. Shchukin, and D.~E. Jesson, Phys. Rev. B {\bf 72}, 245332 (2005).

\bibitem{hammerschmidt}
T. Hammerschmidt, P. Kratzer, and M. Scheffler, unpublished.

\bibitem{pryor}
C. Pryor, J. Kim, L.~W. Wang, A.~J. Williamson, and A. Zunger, J. Appl. Phys. {\bf 83}, 2548 (1998).

\bibitem{bester_a}
G. Bester, X. Wu, D. Vanderbilt, and A. Zunger, Phys. Rev. Lett. {\bf 96}, 187602 (2006).

\bibitem{bester_b}
G. Bester, A. Zunger, X. Wu, and D. Vanderbilt, Phys. Rev. B {\bf 74}, 081305 (2006).

\bibitem{rodt05}
S. Rodt, A. Schliwa, K. P\"otschke, F. Guffarth, and D. Bimberg, Phys. Rev. B {\bf 71}, 155325 (2005).

\end{thebibliography}
\end{document}